\documentclass[aps,pra,showpacs]{revtex4}
\usepackage{graphicx}

\begin{document}

\title{Symmetrization and Entanglement of Arbitrary States of Qubits}
\author{M. Asoudeh}\email{masoudeh@mehr.sharif.edu}
\author{V. Karimipour } \email{vahid@sharif.edu}
\author{L. Memarzadeh}\email{laleh_memarzadeh@mehr.sharif.edu}
\author{A. T. Rezakhani } \email{tayefehr@mehr.sharif.edu}
\affiliation{Department of Physics, Sharif University of
Technology, P.O. Box 11365-9161, Tehran, Iran}
\date{\today}

\begin{abstract}

Given two arbitrary pure states $ |\phi\rangle $ and $
|\psi\rangle $ of qubits or higher level states, we provide
arguments in favor of states of the form $
\frac{1}{\sqrt{2}}(|\psi\rangle |\phi\rangle + i |\phi\rangle
|\psi\rangle ) $ instead of symmetric or anti-symmetric states,
as natural candidates for optimally entangled states constructed
from these states. We show that such states, firstly have on the
average a high value of concurrence, secondly can be constructed
by a universal unitary operator independent of the input states.
We also show that these states are the only ones which can be
produced with perfect fidelity, by any quantum operation designed
for intertwining two pure states with a relative phase. A
probabilistic method is proposed for producing any pre-determined
relative phase into the combination of any two arbitrary states.
\end{abstract}
\maketitle
\section{Introduction}
Entanglement \cite{epr,sch} is a quantum mechanical resource that
can be used for many computational and communication purposes.
During the past few years many experimental efforts have been
reported for creating entanglement \cite{exp} along with
theoretical scenarios \cite{zanardi, dur, cirac, krauscirac, bhe,
vers} for generating as much entanglement as possible. A
remarkable scenario has been put forward by Buzek and Hillery in
\cite{bhe} which stems from the most apparent property of
entangled states, that is, their symmetry property. Consider two
systems $A$ and $B$ in a pure state $|\Psi\rangle_{AB}$ which can
not be decomposed into a product of state vectors of the two
parts. Such a state is called entangled and has a value of
entanglement according to various measures of entanglement,
defined for measuring this property. A symmetric state of the
form $|\Psi\rangle_{AB} = N(|\psi\rangle_A \otimes |\phi\rangle_B
+ |\phi\rangle_A \otimes |\psi\rangle_B)$ is a
prototype of a pure state having this property.\\
The scenario of \cite{bhe} is thus based on the following natural
question: Given two systems $A$ and $B$ respectively in pure
states $|\psi\rangle$ and $|\phi\rangle$ ,is it possible to
construct a quantum machine $M$ which takes these two states as
input and produces with exact fidelity, a symmetric and hence
entangled output state?  An even simpler task for this machine is
to symmetrize an unknown state $|\psi\rangle$ of $A$ with a fixed
reference state $|\phi\rangle $ of $B$, that is
\begin{equation}\label{first}
 |\psi\rangle_A \otimes|\phi\rangle_B\otimes |v_0\rangle_M \longrightarrow
 (|\psi\rangle_A\otimes|\phi\rangle_B + |\phi\rangle_A\otimes |\psi\rangle_B)
 \otimes|v_{\psi}\rangle_M
\end{equation}
where $|v_0\rangle $ and $ |v_{\psi}\rangle$ are the initial and
final states of the machine. The authors of \cite{bhe} show by a
simple argument that linearity and unitarity of quantum mechanics
do not allow such machines to exist. Remarkably however, they
succeed to construct an optimal machine which produces an output
mixed state $\rho^{{\rm{out}}}_{AB}$ which has a universal
(input-independent) fidelity equal to $ \frac{9+3\sqrt{2}}{14} =
0.946$, with the ideal symmetric state $|\Psi\rangle_{AB} =
N(|\psi\rangle_A \otimes |\phi\rangle_B + |\phi\rangle_A \otimes
|\psi\rangle_B)$. Here $N$ is a normalization constant. They then
proceed to show that the output impure state
$\rho^{{\rm{out}}}_{AB}$ is indeed quantum mechanically entangled
or inseparable by using the Peres-Horodecki's criterion
\cite{peres, horo} and showing that for all input states the
partially transposed matrix $(\rho_{AB})^{T_{A}}$ has one
negative eigenvalue. The negativity of this eigenvalue
however depends on the input state.\\
We should also mention another related scheme for optimal
entangling, the one proposed by Alber \cite{alber} who considers
anti-symmetrization of an arbitrary input state with a reference
state yielding the result that the output is a maximally
disordered mixture of anti-symmetric Bell states. For the case of
qubits however, there is only one anti-symmetric Bell state,
namely $|\phi^-\rangle := \frac{1}{\sqrt{2}}(|01\rangle -
|10\rangle)$ so that the output carries
no information about the input states.\\

In this paper we want to study another scenario for entangling
states. Our scenario is based on the idea that optimal
entanglement need not necessarily be obtained by symmetrization.
By looking at symmetrization and entanglement as two different
tasks we propose different methods for their production.  Our
scenario may not be optimal in the class of all conceivable
operations, however it is simple both theoretically and
experimentally. Moreover it clarifies to some extent the relation
between symmetrization and entanglement. By discussing it we hope
at least to raise some
questions for further study.\\
The paper is organized as follows: In Sec.~\ref{sec2}, we propose
a simple method for entangling qubits based on maximization of the
entanglement of formation of a mixed state for which we have a
closed formula as given by Wootters \cite{wootters}. In
Sec.~\ref{sec3}, we consider the problem of symmetrization
separately and show that given two arbitrary states $|\phi\rangle
$ and $ |\psi\rangle $ no quantum operation can produce a state
of the form
\begin{equation}\label{new}
 N(|\phi\rangle\otimes |\psi\rangle + e^{i\theta} |\psi\rangle
\otimes|\phi\rangle),
\end{equation}
 unless $\theta = \pm
\frac{\pi}{2}.$ Thus the only machine which can produce with
exact fidelity, linear combinations of the above form for
arbitrary states, is the one found in Sec.~\ref{sec2} for optimal
entanglement. In this same section we propose a probabilistic
method for producing generalized symmetric states of two
arbitrary states, i.e. states of the form given in (\ref{new}),
where $ N$ is a normalization constant and $ e^{i\theta}$ is any
predetermined phase.
\section{Production of maximal entanglement}\label{sec2}
Suppose that we have two qubits $A$ and $B$ in arbitrary pure
states $| \psi\rangle_A $ and $ |\phi\rangle_B $. When we want to
produce entangled states of the form similar to the one discussed
in the introduction (i.e. symmetric or anti-symmetric states), by
passing these qubits from a quantum machine $M$, it is rather
natural to think that an obstacle is that at the output the two
parts are entangled with the machine itself in the form
\begin{equation}\label{1}
  |\Psi^{{\rm{out}}}\rangle := |\psi\rangle_A |\phi\rangle_B |X\rangle_M + |\phi\rangle_A|\psi\rangle_B |Y\rangle_M.
\end{equation}
where $|X\rangle $ and $|Y\rangle$ are states of the machine. It
is easy to see that such an operation is linear and unitary
provided that the machine states be two fixed input-independent
states satisfying the following relations
\begin{eqnarray}\label{2}
  &\langle X|X\rangle + \langle Y|Y\rangle = 1 \hskip 2cm \langle X|Y\rangle + \langle Y|X\rangle = 0
\end{eqnarray}
A convenient parametrization of these inner products is
\begin{eqnarray}\label{3}
 &\langle X|X\rangle = \frac{1+\xi}{2}, \ \ \ \ \langle Y|Y\rangle = \frac{1-\xi}{2}, \ \ \ \ \langle X|Y\rangle = i\eta
\end{eqnarray}
where $\xi$ and $ \eta $ are real parameters.  We now ask under
what condition, the output density matrix
$\rho^{{\rm{out}}}_{AB}$ of the two systems $A$
and $ B $ has the maximum value of entanglement.\\
For two qubits in a mixed state $\rho$, we have a well
established measure of entanglement given by a closed formula. It
has been introduced by Wootters and Hill in \cite{woottershill,
wootters} and is directly related to the entanglement of
formation of such a mixed state. It is called concurrence and is
denoted by $C$ and is given by
\begin{eqnarray}\label{4}
  C = {\text{max}}~ (0, \lambda_1-\lambda_2-\lambda_3-\lambda_4)
\end{eqnarray}
where $ \lambda_1 $ to $ \lambda_4 $ are the eigenvalues of the
following matrix in decreasing order:
\begin{equation}\label{R}
  R = \sqrt{\rho^{\frac{1}{2}}\tilde{\rho}\rho^{\frac{1}{2}}}.
\end{equation}
Here
\begin{equation}\label{R1}
  \tilde{\rho} = (\sigma_2\otimes \sigma_2)\rho^* (\sigma_2\otimes
  \sigma_2),
\end{equation}
$\sigma_2 $ is the second Pauli matrix and $\rho^* $ is  the
complex conjugate of $\rho $ in the computational basis. Equivalently $\lambda_i $'s can be taken to be the
square root of eigenvalues of the matrix $ \rho \tilde{\rho}$. \\
Here and in what follows we designate the density matrix of the
two qubits simply by $ \rho $ instead of $ \rho_{AB}$.  For the
calculation of eigenvalues we can take without loss of generality,
$|\psi\rangle = a |0\rangle + b|1\rangle $ and $ |\phi\rangle =
|0\rangle $ to obtain
\begin{equation}\label{output}
  |\Psi^{{\rm{out}}}\rangle = \left(\begin{array}{c}
    a|X\rangle + a|Y\rangle \\
    b|Y\rangle \\
    b|X\rangle \\
    0 \
  \end{array}\right)
\end{equation}
from which we obtain $\rho^{{\rm{out}}}$ by taking the trace of
$|\Psi^{{\rm{out}}}\rangle \langle \Psi^{{\rm{out}}}|$ over the
machine states. The result is
\begin{eqnarray}\label{6}
\rho = \left( \begin{array}{cccc}
  |a|^2 & a\overline{b}(-i\eta + \frac{1-\xi}{2}) & a\overline{b}(i\eta + \frac{1+\xi}{2}) & 0 \\
  \overline{a}b(i\eta + \frac{1-\xi}{2}) & |b|^2(\frac{1-\xi}{2}) & i|b|^2 \eta & 0 \\
  \overline{a}b(-i\eta + \frac{1+\xi}{2}) & -i|b|^2\eta & |b|^2(\frac{1+\xi}{2}) & 0 \\
  0 & 0 & 0 & 0
\end{array}\right)
\end{eqnarray}
where we have used the relations in (\ref{2}).\\
Using (\ref{R1}) we obtain
\begin{eqnarray}\label{7}
\tilde{\rho} = \left(
\begin{array}{cccc}
  0 & 0 & 0 & 0 \\
  0 & |b|^2(\frac{1+\xi}{2}) & i|b|^2\eta & -a\overline{b}(i\eta + \frac{1+\xi}{2}) \\
  0 & -i|b|^2\eta & |b|^2(\frac{1-\xi}{2}) & -a\overline{b}(-i\eta + \frac{1-\xi}{2}) \\
  0 & -\overline{a}b(-i\eta + \frac{1+\xi}{2}) & -\overline{a}b(i\eta + \frac{1-\xi}{2}) & |a|^2
\end{array}
\right).
\end{eqnarray}

Finally the matrix $ \rho \tilde{\rho}$ is found to be

\begin{equation}\label{rhorho}
\rho\tilde{\rho} = \left(
\begin{array}{cccc}
  0 & a\overline{b}|b|^2m_- & a\overline{b}|b|^2m_+
  & -2(a\overline{b})^2(\eta^2+\frac{1-\xi^2}{4}-i\eta \xi) \\
  0 & |b|^4(\eta^2+\frac{1-\xi^2}{4}) & i|b|^4\eta(1-\xi) & -a\overline{b}|b|^2m_+ \\
  0 & -i|b|^4\eta(1+\xi) & |b|^4(\eta^2+\frac{1-\xi^2}{4}) & -a\overline{b}|b|^2m_- \\
  0 & 0&0 &0
\end{array}
\right),
\end{equation}
where $ m_{\pm} := (\eta^2+\frac{1-\xi^2}{4}\pm i\eta(1\mp \xi))$.

The eigenvalues of this matrix are easily determined, it has
obviously two zero eigenvalues, the other two being the
eigenvalues of the central 2 by 2 sub-matrix. The square root of
its eigenvalues which are the eigenvalues of the matrix $ R $ are
in decreasing order
\begin{eqnarray}\label{8}
 &\lambda_1 = |b|^2(\frac{\sqrt{1-\xi^2}}{2}+\eta)\ \ \ \
 &\lambda_2 = |b|^2(\frac{\sqrt{1-\xi^2}}{2}-\eta)\ \ \ \lambda_3 = 0 \ \ \ \lambda_4=0.
\end{eqnarray}
where in taking the square roots we have used the Chauchy
Schwartz inequality for the inner products in (\ref{3}) which
implies that $ \eta \leq \frac{\sqrt{1-\xi^2}}{2}$.\\
Putting all this together with equation (\ref{4}), leads to the
following value for the concurrence

\begin{eqnarray}\label{9} &C =
2 |b|^2 \eta .
\end{eqnarray}
To obtain maximum concurrence we have to take $ \eta =
\frac{1}{2} $ which according to the Cauchy Schwartz inequality
forces us to choose $ \xi = 0 $.  This then means that the
vectors $|X\rangle $ and $ |Y\rangle $ are the same modulo a
crucial phase, that is
\begin{eqnarray}\label{10}
&|X\rangle = \frac{1}{\sqrt{2}} |e\rangle \ \ \ \ \ |Y\rangle =
\frac{i}{\sqrt{2}} |e\rangle,
\end{eqnarray}
where $ |e\rangle $ is a normalized state. Interestingly the
output state now turns out to be disentangled from the machine,
so that the machine produces a pure entangled state:
\begin{equation}\label{11}
 |\psi\rangle_A \otimes |\phi\rangle_B \longrightarrow |\Psi\rangle^{{\rm{out}}}_{AB} :=
  \frac{1}{\sqrt{2}}(|\psi\rangle_A\otimes |\phi\rangle_B + i|\phi\rangle_A \otimes|\psi\rangle_B)
\end{equation}
Having determined the optimal choice of $\eta $ by taking the
state $|\phi\rangle $ in the  basis state $ |0\rangle $, we can
now calculate the concurrence produced for any two arbitrary
input states. Since the output state is pure, its concurrence can
be calculated from an alternative formula for any two input pure
states
\begin{eqnarray}\label{12}
  &C = |\langle\Psi|\tilde{\Psi}\rangle|
\end{eqnarray}
where we have abbreviated $|\Psi^{{\rm{out}}}_{AB}\rangle $ to
$|\Psi\rangle $ and
\begin{eqnarray}\label{13}
&|\tilde{\Psi}\rangle = (\sigma_2 \otimes \sigma_2) |\Psi^*\rangle
=
  \frac{1}{\sqrt{2}}(|\tilde{\psi}\rangle_A \otimes|\tilde{\phi}\rangle_B -
  i|\tilde{\phi}\rangle _A\otimes|\tilde{\psi}\rangle_B).
\end{eqnarray}
Inserting this in (\ref{12}) and using the fact that for every
qubit state $ |\psi\rangle $, $\langle \psi|\tilde{\psi}\rangle =
0 $, we obtain
\begin{eqnarray}\label{14}
 & C = \langle \psi|\tilde{\phi}\rangle\langle \phi|\tilde{\psi}\rangle
\end{eqnarray}
If the two initial states are two spin states in definite
directions on the Bloch sphere, that is if $ |\psi\rangle = |
{\bf{\hat{n}}}\rangle $ and $ |\phi\rangle =
|{\bf{\hat{m}}}\rangle $ we find after straightforward
calculations that
\begin{eqnarray}\label{15}
  &C = \frac{1}{2}(1- \bf{\hat{n}\cdot \hat{m}}).
\end{eqnarray}
The details of this calculation is given in the appendix. The
average of this concurrence over all input states is
$\frac{1}{2}$. Thus this transformation produces on the average,
an entanglement which as measured by concurrence is $1/2$.
Equation (\ref{15}) also shows that maximal entanglement is
produced by intertwining two anti-parallel spin states on the
Bloch sphere.\\
Up to now we have shown the possibility of the desirable
entangling transformation. In view of (\ref{11}) the actual form
of the transformation is given by
\begin{eqnarray}\label{5}
&U=\frac{1}{\sqrt{2}}(I + i P) = e^{i\frac{\pi}{4}P} =
e^{i\frac{\pi}{8}}e^{i\frac{\pi}{8}{\bf{\overrightarrow{\sigma}
\cdot \overrightarrow{\sigma}}}}
\end{eqnarray}
where $ P$ is the permutation operator (i.e. $P|\alpha,
\beta\rangle = |\beta, \alpha\rangle $ ) which for two
dimensional spaces is related to the Pauli matrices as $ P =
\frac{1}{2}(I+ {\bf {\overrightarrow{\sigma} \cdot
\overrightarrow{\sigma}}}) $. The above argument and the
combination of the last two formulas also tell us how to produce
states with a definite amount of entanglement or concurrence,
when we have partial information about the input states. Let us
fix the spin state $|\phi\rangle = |{\bf{\hat{m}}}\rangle $ in the
direction $z$. We then need only take a spin state $|\psi\rangle =
|{\bf{\hat{n}}}\rangle $ with ${\bf{\hat{n}}}$ making an angle
$\theta$ with the $z$ axis (this still leaves the angle $\phi $
undetermined) and apply the spin spin interaction (\ref{5}) to
produce a concurrence of $ C = \frac{1}{2}(1- \cos{\theta})$. In
this way by just adjusting the initial value of $ \theta$ we can
produce entangled states with any desired value of entanglement,
ranging from the minimum value of $ C = 0 $ for $ \theta = 0 $ to
the
maximum value of $ C = 1 $ for $ \theta = \pi $.\\
Moreover the output density matrix also retain some information
about the input states, since it is easily verified from
(\ref{11}) that
\begin{eqnarray}\label{rho}
 & \rho^{\rm{(out)}}_{A} =  tr_B(|\Psi^{{\rm{out}}}\rangle_{AB}\langle \Psi^{{\rm{out}}}|)
 = \frac{1}{2}(|\psi\rangle_A\langle \psi| + |\phi\rangle_A\langle
 \phi|),
\end{eqnarray}
with an identical formula for $ \rho^{{\rm{out}}}_B $. Thus the
fidelity of reduced one particle density matrices of the output
with any of the states $ |\psi\rangle $ or
$|\phi\rangle $ is $ \frac{1}{2}$.\\
If we are interested in determining how much the output state $
|\Psi^{\rm{out}}\rangle = \frac{1}{\sqrt{2}}(|\psi\rangle
|\phi\rangle + i |\phi\rangle |\psi\rangle)$ is close to a
symmetric state $ |\Psi^{\rm{sym}}\rangle = N(|\psi\rangle
|\phi\rangle + |\phi\rangle |\psi\rangle)$, then we can find the
overlap of these two states. Knowing that $ 2N^2(1+|\langle
\psi|\phi\rangle|^2)=1$, we find:
\begin{eqnarray}\label{overlap}
  &|\langle \Psi^{\rm{(out)}}|\Psi^{\rm{sym}}\rangle|^2 =
  \frac{1}{2}(1 + |\langle \psi|\phi\rangle|^2 ).
\end{eqnarray}
When averaged over all the input states this will give an overlap
of $\frac{3}{4}= 0.75 $.
\section{Symmetrization}\label{sec3}
\begin{figure}[tp]
\includegraphics[width=13.5cm,height=3cm]{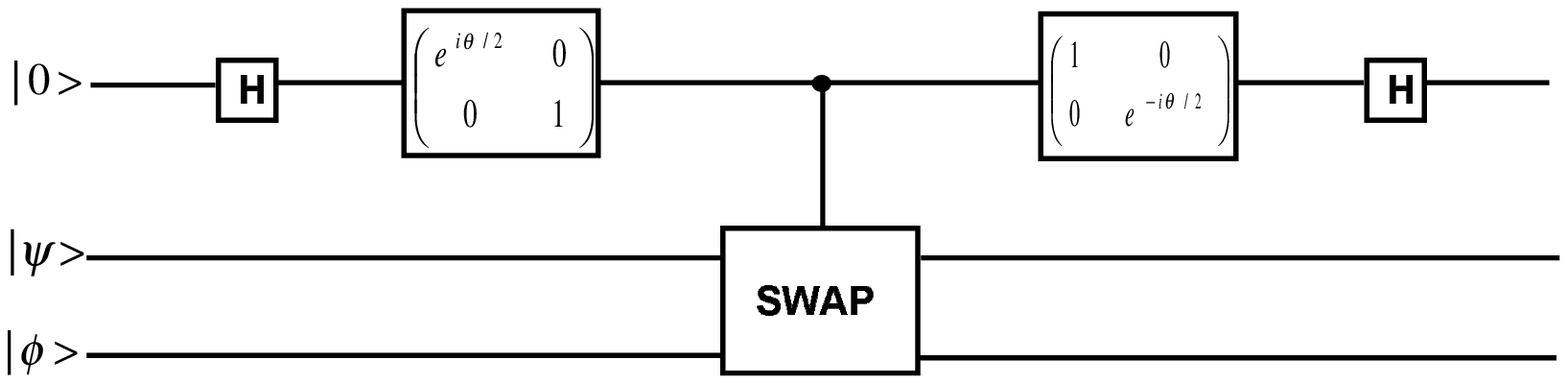}
\caption{Quantum network for performing probabilistic generalized
symmetrization. } \label{fig.1}
\end{figure}

By separating the issue of symmetrization from that of
entanglement we will have more freedom in constructing states
exhibiting each of these properties. Concerning symmetrization
problem, we can now
ask a more general question than the one considered in \cite{bhe}.\\
 Is it possible to have a quantum machine which takes two input
 states $ |\psi\rangle $ and $ |\phi\rangle $ and produces generalized symmetric state as follows?
\begin{equation}\label{machine}
|\psi\rangle_A|\phi\rangle_B |v\rangle_M \longrightarrow
(|\psi\rangle|\phi\rangle + e^{i\theta} |\phi\rangle
|\psi\rangle)_{AB}
 |v'\rangle_M .\\
\end{equation}
where $ \theta $ is a predetermined phase. Note that for
simplicity we have suppressed all the $ \otimes $ signs. Here $
|v\rangle $ is the initial normalized state of the machine and $
|v'\rangle $ is the state of the machine after operation and it
certainly depends on the initial states $|\psi\rangle $ and $
|\phi\rangle $.\\ We will see that such a machine exists only
when $ e^{i\theta} = i $. To see this we note that if such a
machine exists, it has to act as follows
\begin{eqnarray}\label{bh1}
 |0\rangle |0\rangle |v\rangle &\longrightarrow & |0\rangle
  |0\rangle |v_0\rangle\\
  |1\rangle |0\rangle |v\rangle &\longrightarrow &(|1\rangle
  |0\rangle + e^{i\theta} |0\rangle |1\rangle) |v_1\rangle.
\end{eqnarray}
where $|v_0\rangle $ and $ |v_1\rangle $ are two of the machine
states. Unitarity then demands the following relations among these
machine states
\begin{eqnarray}\label{bh2}
&\langle v_0|v_0\rangle = 1\hskip 2cm \langle v_1|v_1\rangle =
\frac{1}{2}.
\end{eqnarray}
We now consider an input state like $ \frac{1}{\sqrt{2}}
(|0\rangle + |1\rangle )|0\rangle $ which according to
(\ref{machine})
 should be transformed to
\begin{equation}\label{bh3}
 |\Psi\rangle^{\rm{id}} = \frac{1}{\sqrt{2}} \Big(|0\rangle|0\rangle + |1\rangle|0\rangle + e^{i\theta}
 (|0\rangle|0\rangle+|0\rangle|1\rangle)\Big) |v_2\rangle
\end{equation}
where $|v_2\rangle $ is another state of the machine. On the other
hand linearity of quantum mechanics requires that in view of
(\ref{bh1}), this input state be transformed to the state
\begin{eqnarray}\label{bh4}
 |\Psi\rangle^{\rm{out}}=\frac{1}{\sqrt{2}} \Big(|0\rangle |0\rangle |v_0\rangle +
  (|1\rangle |0\rangle + e^{i\theta}|0\rangle
  |1\rangle ) |v_1\rangle\Big)
  . \end{eqnarray}
Comparing $|\Psi^{\rm{id}}\rangle $ and $|\Psi^{\rm{out}}\rangle
$, we find $ |v_0\rangle = (1+e^{i\theta})|v_2\rangle $ and $
|v_2\rangle = |v_1\rangle$, which are compatible with the norm
condition (\ref{bh2}) only if $e^{i\theta} = i$. This proves the
negative part of the theorem. The positive part has been
demonstrated already in section 2, where we have shown that the
operator $U=e^{i\frac{\pi}{4}P}$ intertwines any two arbitrary
states $|\psi\rangle $ and $ |\phi\rangle $ to
$\frac{1}{\sqrt{2}}(|\psi\rangle |\phi\rangle + i |\phi\rangle
|\psi\rangle $.\\
The interesting point is that from two different starting points
we have arrived at the above form of states, one from requiring
maximum entanglement production in section 2 and the other from
requiring universal symmetrization. Note that while the arguments
and the results for entanglement production are specific to
qubits, those for symmetrization are valid in any dimension. This
raises the question as to whether this coincidence holds true
also in other dimensions. That is if the states of the form
$\frac{1}{\sqrt{2}}(|\psi\rangle |\phi\rangle + i |\phi\rangle
|\psi\rangle $ are good candidates for highly entangled states in
higher dimensions? We will touch upon this question in the
conclusion of the paper. \\
Although we do not have a quantum machine which can put two
arbitrary states $ |\psi\rangle $ and $ |\phi\rangle $ into a
combination with a general relative phase $ e^{i\theta}\ne i $ ,
we can achieve this probabilistically by using Fredkin gates which
do controlled swap operations. The circuit shown in
Fig.~\ref{fig.1} which is a generalization of the one given in
\cite{barenco} for producing symmetric and antisymmetric
superposition of input states, performs such an operation. Note
that no three body interaction is needed for implementing the
Fredkin gate and such a gate can be constructed by a combination
of two body operations exactly \cite{wilcek}. In fact such a
circuit develops the state $ |\psi\rangle |\phi\rangle |0\rangle $
to the state
\begin{eqnarray}\label{generalizedsymmetrizer}
&|\Psi^{{\rm{out}}}\rangle =
\frac{1}{2}(e^{\frac{-i\theta}{2}}|\phi\rangle|\psi\rangle +
e^{\frac{i\theta}{2}}|\psi\phi\rangle)_{AB}|0\rangle_M
+\frac{1}{2}(e^{\frac{i\theta}{2}}|\psi\phi\rangle -
e^{\frac{-i\theta}{2}}|\phi\psi\rangle)_{AB}|1\rangle_M
\end{eqnarray}

 Before
measurement of the control qubit the output state of $A$ and $B$
is a separable and hence disentangled mixed state  given by the
density matrix
\begin{eqnarray}\label{separable}
 & \rho^{\rm{(out)}}_{AB} = \frac{1}{2}\Big( |\psi\rangle \langle \psi| \otimes
  |\phi\rangle \langle \phi|+|\phi\rangle \langle \phi|\otimes |\psi\rangle \langle
  \psi|\Big).
\end{eqnarray}
Once the control qubit (the state of the machine) is measured in
the basis $|0\rangle $ and $|1\rangle $, the two systems $ A$ and
$B$ will be projected onto one of the normalized states
\begin{eqnarray}\label{states}
|\Psi_+\rangle &=& \frac{1}{N_+}(|\phi\psi\rangle
+e^{i\theta}|\psi\phi\rangle)\\
|\Psi_-\rangle &=& \frac{1}{N_-}(|\phi\psi\rangle -e^{i\theta}
|\psi\phi\rangle)
\end{eqnarray}
where $ N_+$ and $N_-$ are normalization factors.  The result of
the measurement of the control bit will declare which one of
these two states have been produced. These states are produced
with probabilities $ P_+ = \frac{N_+^2}{4}$ and $ P_- =
\frac{N_-^2}{4}$. These states have now quantum entanglement as
measured by their concurrence given by
\begin{eqnarray}\label{concurrence}
C_+ &=& |\langle \Psi_+|\tilde{\Psi_+}\rangle| =
\frac{2}{N_+^2}\langle
\phi|\tilde{\psi}\rangle\langle\psi|\tilde{\phi}\rangle\\
C_- &=& |\langle \Psi_-|\tilde{\Psi_-}\rangle| =
\frac{2}{N_-^2}\langle
\phi|\tilde{\psi}\rangle\langle\psi|\tilde{\phi}\rangle
\end{eqnarray}
The average concurrence of the states produced will be given by
\begin{equation}\label{average}
  \overline{C} = P_+ C_+ + P_- C_- =
  \langle \phi|\tilde{\psi}\rangle\langle
  \psi|\tilde{\phi}\rangle,
\end{equation}
which is equal to the concurrence we obtain for the state $
\frac{1}{\sqrt{2}}(|\phi\rangle |\psi\rangle + i|\psi\rangle
|\phi\rangle)$. \\
However as far as entanglement production is concerned, this
machine has no advantage over the simple deterministic machine
proposed in section II, since with that transformation we could
exactly produce every desirable value including the maximum value
of
entanglement.\\

\section{Discussion}\label{sec4}
We have shown that there is no quantum mechanical process which
can change a product state $ |\psi\rangle |\phi\rangle $ into a
state $N(|\psi\rangle |\phi\rangle + e^{i\theta} |\phi\rangle
|\psi\rangle) $ with perfect fidelity unless $e^{i\theta}=\pm i$.
This result is true in any dimension. In two dimensions where we
are dealing with qubits and have a closed formula for the
entanglement of formation of a mixed state \cite{wootters,
woottershill}, we have shown that these states are also the ones
which have the maximum value of entanglement, when $|\phi\rangle
$ and $ |\psi\rangle $ are two initially fixed states. An
interesting question is whether this correspondence exists also
in higher dimensions or not. If this correspondence holds also in
higher dimension then we can conjecture that the states of the
form $|\Psi\rangle = \frac{1}{\sqrt{2}}(|\psi\rangle |\phi\rangle
+ i |\phi\rangle |\psi\rangle) $ rather the symmetric or
anti-symmetric ones are the states which have the maximum
entanglement when averaged over all the input product states. A
hint already comes from calculating the I-concurrence of these
states as given in \cite{runga}. The I-concurrence of a pure
state $|\Psi\rangle_{AB} $ of two systems $A$ and $B$ is defined
to be
\begin{equation}\label{I}
  C = \sqrt{1-2tr (\rho_A)^2} = \sqrt{1-2tr (\rho_B)^2}
\end{equation}
where $ \rho_A $ and $ \rho_B$ are the reduced density matrices
of the subsystems $A$ and $ B $ respectively. If the two systems
are in a pure state as above (i.e. $|\Psi\rangle_{AB}
=\frac{1}{\sqrt{2}}(|\psi\rangle |\phi\rangle + i |\phi\rangle
|\psi\rangle)_{AB} $) we find that
\begin{equation}\label{II}
  \rho = \frac{1}{2}(|\phi\rangle \langle \phi| + |\psi\rangle \langle \psi|
  + i \langle \phi|\psi\rangle |\phi\rangle \langle \psi| - i \langle \psi|\phi\rangle |\psi\rangle \langle
  \phi|),
\end{equation}
where $ \rho $ stands for the density matrix of any of the two
subsystems. A simple calculation shows that
\begin{equation}\label{III}
  C = 1 - |\langle \phi|\psi\rangle|^2
\end{equation}
This concurrence reduces to (\ref{15}) for qubits calculated from
Wootters formula. Moreover for orthogonal states it gives the
maximum value $1$ and when averaged over all product states it
gives the value $ \frac{1}{2} $.
\section{Appendix}
In this section we present in detail the calculation leading to
equation (\ref{15}). For two qubits $|\psi\rangle = \left(\begin{array}{c}
  a \\
  b
\end{array}\right) $ and $|\phi\rangle = \left(\begin{array}{c}
  c \\
  d
\end{array}\right) $, we have according to (\ref{14}), $ C = |ad-bc|^2 $.
Let the qubit states $|\psi\rangle $ and $ |\phi\rangle $
correspond to unit vectors $ {\bf{n}}= (\sin \theta \cos \phi,
\sin \theta \sin \phi,\cos \theta) $ and ${\bf{m}} = (\sin
\theta' \cos \phi', \sin \theta' \sin \phi',\cos \theta') $
respectively. Then we will have
\begin{equation}\label{app1}
  |\psi\rangle \equiv |{\bf{n}}\rangle =\left(\begin{array}{c}
  \cos \frac{\theta}{2} e^{-i\frac{\phi}{2}} \\
  \sin \frac{\theta}{2} e^{i\frac{\phi}{2}}
\end{array}\right)\hskip 2cm |\phi\rangle \equiv |{\bf{m}}\rangle =\left(\begin{array}{c}
  \cos \frac{\theta'}{2} e^{-i\frac{\phi'}{2}} \\
  \sin \frac{\theta'}{2} e^{i\frac{\phi'}{2}}
\end{array}\right)
\end{equation}
from which we find
\begin{equation}\label{app2}
  C = |\sin \frac{\theta'}{2} \cos \frac{\theta}{2}e^{-i\frac{\phi-\phi'}{2}}-
  \sin \frac{\theta}{2} \cos
  \frac{\theta'}{2}e^{i\frac{\phi-\phi'}{2}}|^2.
\end{equation}
Simplifying this expression leads to
\begin{equation}\label{app3}
  C = \frac{1}{2}(1-\cos \theta \cos \theta' - \sin \theta \sin \theta'\cos(\phi - \phi'))
\end{equation}
which is nothing but the expression $C=\frac{1}{2}(1-{\bf{m\cdot n
}})$ written in components.

\end{document}